\documentclass[%
 reprint,
 amsmath,amssymb,
 aps,
 prl,
]{revtex4-1}

\usepackage{graphicx}
\usepackage{dcolumn}
\usepackage{bm}
\usepackage{hyperref}
\usepackage{color}

\begin{document}

\preprint{APS/123-QED}

\title{Nanoskyrmion engineering with $sp$-electron materials: \\ Sn monolayer on SiC(0001) surface}

\author{Danis I. Badrtdinov$^{1}$, Sergey A. Nikolaev$^{1,2}$, Alexander N. Rudenko$^{3,1,4}$, Mikhail I. Katsnelson$^{1,4}$,  and Vladimir V. Mazurenko$^{1}$}

\affiliation{$^{1}$Theoretical Physics and Applied Mathematics Department, Ural Federal University, 620002 Ekaterinburg, Russia \\
$^{2}$ International Center for Materials Nanoarchitectonics, National Institute for Materials Science, 1-1 Namiki, Tsukuba, Ibaraki 305-0044, Japan \\
$^{3}$ School of Physics and Technology, Wuhan University, Wuhan 430072, China \\
$^{4}$ Institute for Molecules and Materials, Radboud University, Heijendaalseweg 135, 6525 AJ Nijmegen, The Netherlands
}

\date{\today}

\begin{abstract}
Materials with $sp$-magnetism demonstrate strongly nonlocal Coulomb interactions, which opens a way to probe correlations in the regimes not achievable in transition metal compounds. By the example of Sn monolayer on SiC(0001) surface, we show that such systems exhibit unusual but intriguing magnetic properties at the nanoscale. Physically, this is attributed to the presence of a significant ferromagnetic coupling, the so-called direct exchange, which fully compensates ubiquitous antiferromagnetic interactions of the superexchange origin. Having a nonlocal nature, the direct exchange was previously ignored because it cannot be captured within the conventional density functional methods and significantly challenges ground state models earlier proposed for Sn/SiC(0001). Furthermore, heavy adatoms induce strong spin-orbit coupling, which leads to a highly anisotropic form of the spin Hamiltonian, in which the Dzyaloshinskii-Moriya interaction is dominant. The latter is suggested to be responsible for the formation of a nanoskyrmion state at realistic magnetic fields and temperatures. 
\end{abstract}

\maketitle


{\it Introduction.} ---$sp$-magnetism in two-dimensional and quasi-two-dimensional materials is an actively developing field of research, which promises exotic and highly tunable magnetic properties. Among them a large Rashba spin splitting \cite{Rashba1,Rashba2}, which is required for spintronics applications, is observed in Pb monolayer deposited on Ge(111) surface \cite{Yaji}. Another promising example is high-spin graphene-based structures, such as triangulene \cite{triangulene}, that can be used for design of high-density ultrafast quantum nanodevices \cite{Yazyev,IBM}. As was shown theoretically in Ref.~\onlinecite{Katsnelson_1}, $sp$-electron magnets can have much higher Curie temperatures than conventional magnetic semiconductors. However, the experimental study of nanostructures with $sp$ electrons
is still a challenging task. The main complications of using standard techniques, such as spin-polarized scanning tunneling microscopy (SP-STM)~\cite{Wiesendanger1}, come from the delocalized nature of magnetic moments in these systems. Recent STM experiments on hydrogen adatoms on graphene have shown that the spin-polarized state extends over several nanometers away from the hydrogen atoms providing the direct coupling between magnetic moments at long distances~\cite{CHH}. 

On the theory side, there is still no consistent description of the magnetic moment in the $sp$ surface nanosystems. 
A number of calculations based on density functional theory (DFT) \cite{Cho, Glass} predicted the antiferromagnetic ground state with the energy gap in the infrared region. 
However, the intrinsic inability of DFT to describe truly nonlocal interactions makes this method less applicable to $sp$ systems, where magnetization density is not well localized at atomic sites. Particularly, strong overlap between the $sp$-wavefunctions can result in a considerable direct exchange interaction, comparable in magnitude with the conventional antiferromagnetic coupling. In the commonly used local spin density approximation, the effective magnetic field is determined only by magnetization density at the same point, which means the absence of direct exchange interactions~\cite{Katsnelson_2}; in this approximation, the exchange interactions arise only from the nonlocality of kinetic energy~\cite{Katsnelson_3} corresponding to the superexchange.
In this situation, description of $sp$-magnetism at the level of Hubbard-like models appears to be more appropriate as it allows one to overcome the limitations of DFT and advance the understanding of magnetic properties in surface nanostructures. 

Recently, it has been shown that adatom $sp$-electron systems such as Si(111):\{C,Si,Sn,Pb\} demonstrate a spin spiral state at zero magnetic field \cite{Danis} due to strong spin-orbit coupling induced by heavy adatoms. Despite the prediction of a skyrmion state, unrealistically large magnetic fields ($\sim$ 190 T) have been proposed. However, the microscopic analysis of magnetic interactions in these systems allows to propose a promising way to tune the material parameters in order to stabilize skyrmionic states achievable under realistic experimental conditions. Since the critical magnetic field for the skyrmion phase is scaled with the isotropic exchange interaction between nearest neighbors, the variation of magnetic couplings 
could be done by changing the chemical composition of the system. 

Here, we propose the Sn adatom layer on SiC(0001) substrate to be an appealing candidate to observe the skyrmion state in STM experiments. We find that the ferromagnetic direct exchange coupling fully compensates the antiferromagnetic interaction of the superexchange origin. The resulting spin Hamiltonian turns out to be strongly anisotropic, with dominating Dzyaloshinskii-Moriya interactions. Finite temperature and magnetic field spin simulations demonstrate that such situation allows for the formation of highly tunable spin spiral and skyrmion states at realistic experimental conditions.

\begin{figure}[!h]
\includegraphics[width=0.46\textwidth]{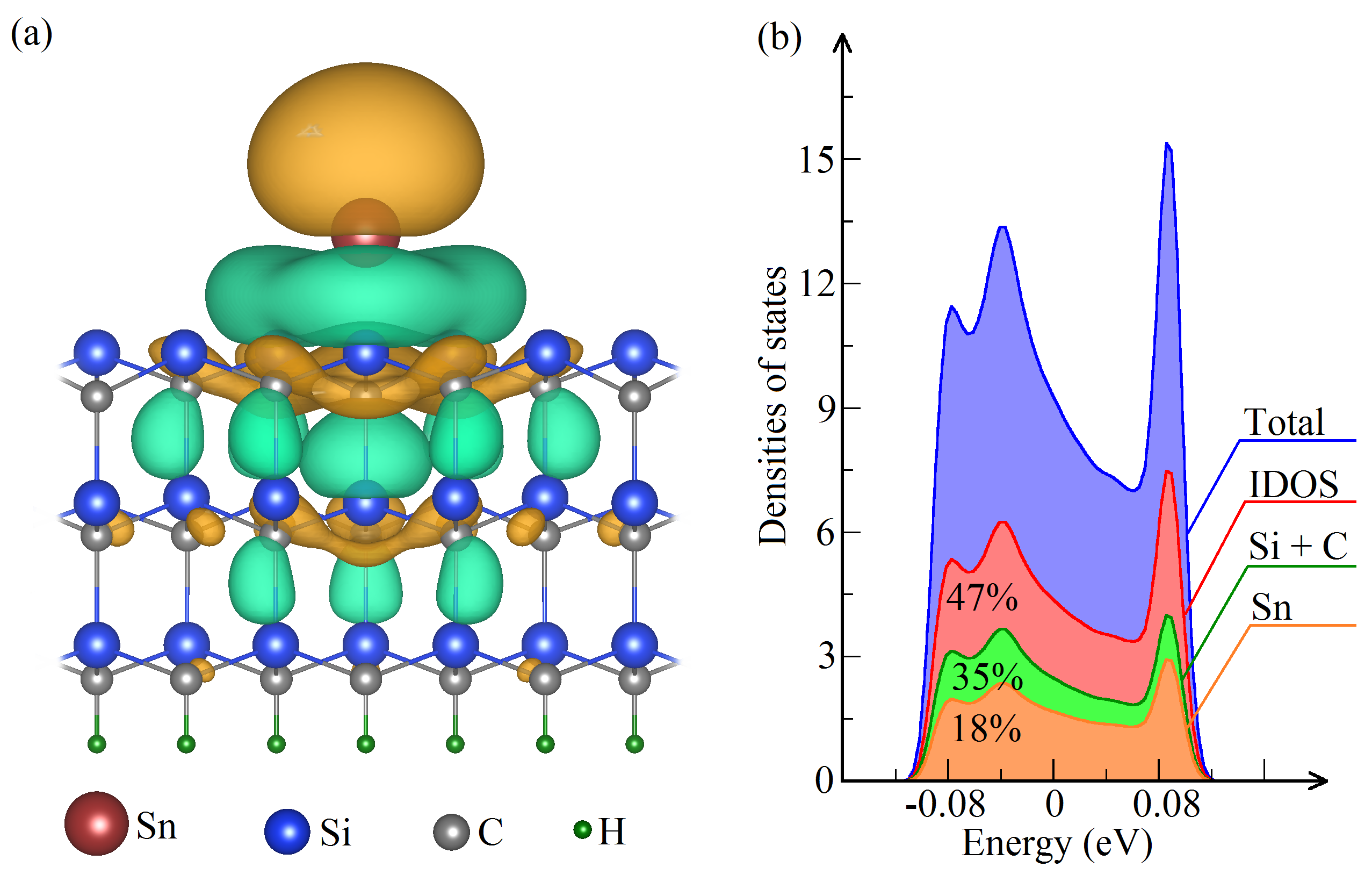}
\caption{ (a) Maximally localized Wannier functions representing the band at the Fermi level in the DFT spectrum and $S=1/2$ state in Sn/SiC(0001). (b) Total and partial densities of states near the Fermi level. IDOS denotes interstitial density of states.}
\label{fig:Crystal}
\end{figure}

{\it Atomic structure and DFT spectrum.} ---The simulated crystal structure of Sn/SiC(0001) system is visualized in Fig.~\ref{fig:Crystal}. It consists of a regular array of Sn atoms placed on top of three silicon carbide layers, and a hydrogen layer used to passivate the lowermost layer of the slab. 
Since the topmost layer of silicon is not passivated, there are uncompensated electrons in Sn/SiC(0001), whose noninteracting energy spectrum is characterized by a half-filled band located at the Fermi level. Given the absence of inversion symmetry (the $C_{3v}$ point group), Kramers degeneracy is lifted resulting in a sizable band splitting at the Fermi energy. The calculation details are presented in Supplemental Material \cite{Supplementary}. The full potential DFT calculations show that the band at the Fermi level is composed of the $p$ states of Sn (18\%), silicon $p$ states (17\%), and carbon $p$ states (8\%). Importantly, there is a large contribution of 47\% coming from the interstitial states, which, as will be shown below, plays an essential role in the formation of magnetic properties in Sn/SiC(0001).

{\it Analysis of noninteracting magnetic solution.} ---Full-potential DFT calculations show that the total magnetization calculated on a minimal ($\sqrt{3}\times\sqrt{3}$) unit cell amounts to 1 $\mu_B$, which corresponds to the $S=1/2$ state realized for the half-filled band. The partial contributions to the total magnetization are related to the composition of the states near the Fermi level. The magnetic moment of the Sn adatom is 0.22 $\mu_B$, while the interstitial contribution to the total magnetization is 0.45 $\mu_B$. To take into account the effect of the interstitial magnetization, it is instructive to describe the magnetic moments in Sn/SiC(0001) using the formalism of Wannier functions. Apart from the advantage of taking into account hybridization effects between different atomic states, this approach allows to quantify the spatial spread of magnetic orbitals and makes it possible to construct a minimal low-energy model, as was done in Ref.~\cite{silke1}. 

Previous DFT studies \cite{Cho, Glass} were mainly focused on searching for the magnetic configuration corresponding to the minimum of the total energy. The collinear AFM configuration was proposed to be the magnetic ground state. However, previous studies did not succeed in constructing a microscopic model that can reproduce ground-state DFT results and did not describe excited states of the system in question. This is the main aim of our work. We analyze the total energies of different magnetic solutions simulated in earlier works \cite{Cho, Glass} by means of the simplest Heisenberg model on the triangular lattice, $\mathcal {H} = \sum_{ij} J_{ij} {\bf S}_{i} {\bf S}_{j}$. To this end, we use the total energy difference method. The details of such estimation are presented in Supplemental Material \cite{Supplementary}. The calculated exchange interactions, 2.53 meV (between nearest neighbors) and 0.45 meV (between next-nearest neighbors) are antiferromagnetic, which originate from the Anderson's superexchange mechanism, $J^{kin}_{ij} = \frac{2t^{2}_{ij}}{U}$, where $t_{ij}$ is the hopping integral and $U$ is the on-site Coulomb interaction. 
Importantly, the kinetic exchange interaction in Sn/SiC(0001) is about five times smaller than that in Si/Si(111), which provides the conditions toward the stabilization of magnetic skyrmions at realistic magnetic fields. The obtained agreement between superexchange theory and total energy difference results indicates that the antiferromagnetic state reported earlier for Sn/SiC(0001) was of the superexchange origin. However, due to spatial delocalization of the magnetic orbital described above, it is important to consider another mechanism arising from the overlap of Wannier functions, namely, the direct exchange.

{\it Direct exchange.} ---Using the calculated Wannier functions, we have estimated the direct exchange interaction introduced by Heisenberg \cite{Heisenberg} to explain the ferromagnetism of iron,
$J^{F}_{ij} = \int \frac{W_{i}^{*}(\boldsymbol{r})W_{j}(\boldsymbol{r})W_{j}^{*}(\boldsymbol{r}')W_{i}(\boldsymbol{r}')}{|\boldsymbol{r}-\boldsymbol{r}'|}  d\boldsymbol{r} d\boldsymbol{r}'$,
where $W_i(\boldsymbol{r})$ is the Wannier function centered on $i$th site.
Being negligibly small in most of the materials with localized magnetic moments, this coupling plays an important role when the antiferromagnetic interactions are suppressed, as it takes place in some copper oxides \cite{LiCu2O2}, where the Wannier functions describing magnetic moments strongly overlap. For Sn/SiC(0001) we found the bare (unscreened) value of the direct exchange interaction to be 5.2 meV, which is considerably larger than the antiferromagnetic exchange of 1.42 meV estimated from superexchange theory~\cite{Supplementary}. Taking into account the effects of screening as well as interactions with next-nearest neighbors does not change the balance between antiferromagnetic and ferromagnetic couplings in the Sn/SiC(0001) system. Therefore, the mean-field ground state of the Sn/SiC(0001) model without spin-orbit coupling is ferromagnetic, which questions previous scenarios for magnetic properties.  

It is worth noting that the standard (semi-)local formulation of DFT does not capture the Heisenberg direct exchange interaction between two different points in space. This is because the variation of the exchange energy splitting on magnetization $\delta E / \delta {\bf m}({\bf r})$ is locally defined, meaning that it is not zero only at the same point ${\bf r}$. To overcome this issue, we construct 
a magnetic model of Sn/SiC(0001) in such a way that it takes into account both the direct exchange interaction and spin-orbit coupling. The latter, as shown in Ref.~\onlinecite{Glass}, strongly affects the noninteracting electronic spectrum near the Fermi level, inducing the splitting of the half-filled band. At the same time, the effect of spin-orbit coupling on magnetic properties of the system remains almost unknown. We found that in view of the compensation of the isotropic exchange interactions, the Dzyaloshinskii-Moriya interaction formed by spin-orbit coupling represents the largest energy scale for magnetic excitations in Sn/SiC(0001).

{\it Anisotropic spin Hamiltonian.} --- The search for correlated systems that are physical realizations of spin Hamiltonians demonstrating unusual magnetic phases is in the focus of condensed matter physics. Exotic magnetic phenomena come into play mostly upon consideration of long-range interactions leading to magnetic frustration, or in the situations when the source of ordinary (ferro-/antiferro-) magnetism, i.e., the isotropic exchange interaction, is suppressed. In the latter case, the excitation spectrum is mainly determined by anisotropic interactions. One remarkable example is $\alpha$-RuCl$_3$ \cite{Kitaev} that is characterized by a strong bond-dependent Kitaev term in the magnetic Hamiltonian. For Sn/SiC(0001), we found that the anisotropic antisymmetric exchange interaction governs magnetic properties of the system. Taking into account the compensation of the ferromagnetic and antiferromagnetic contributions to the isotropic exchange interaction in Sn/SiC(0001), it is instructive to construct and study strongly anisotropic form of the spin Hamiltonian.  

\begin{figure}[t]
\includegraphics[width=0.30\textwidth]{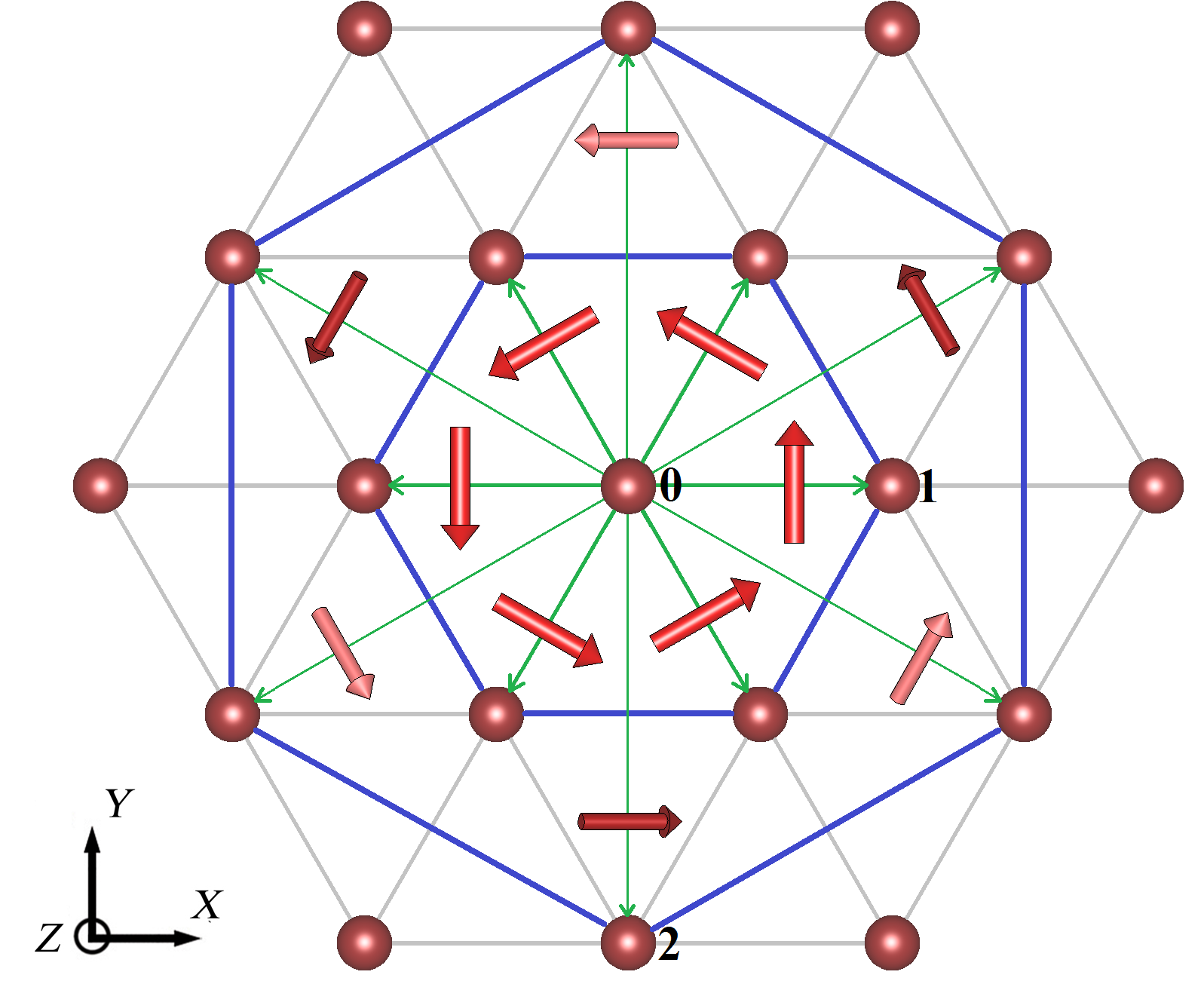}
\caption{ Magnetic model for Sn/SiC(0001). The interaction paths are shown by green lines. Dark and light red arrows denote the DMI vectors for next-nearest neighbor interactions with negative and positive $z$ components, respectively.}
\label{fig:Model}
\end{figure}

We now turn to the low-energy Hubbard model taking into account direct exchange, spin-orbit coupling, and non-local Coulomb interactions \cite{Supplementary}. The solution of the electronic Hamiltonian within the Hartree-Fock approximation results in the opening of an energy gap of 2 eV, in agreement with experiment \cite{Glass}. Further, we consider an extended spin model for Sn/SiC(0001):
\begin{eqnarray}
\hat {\mathcal{H}}^{spin} = \sum_{ij} {\bf D}_{ij}  [\hat {{\bf S}}_{i} \times  \hat {{\bf S}}_{j}] + \sum\limits_{ij}\hat{{\bf S}}_i\overset{\leftrightarrow}{\Gamma}_{ij}\hat{{\bf S}}_j + \sum_{ij} J_{ij} \hat {{\bf S}}_{i} \hat {{\bf S}}_{j}, 
\label{spinham}
\end{eqnarray}
and estimate the corresponding parameters by using superexchange theory. In Eq.~(\ref{spinham}), $\hat {\bf S}$ is the spin operator, $J_{ij}$, ${\bf D}_{ij}$ and $\overset{\leftrightarrow}{\Gamma}_{ij}$ are the isotropic exchange coupling, antisymmetric anisotropic (Dzyaloshinskii-Moriya), and symmetric anisotropic interactions, respectively. 
The symmetry of the obtained Dzyaloshisnkii-Moriya interactions is visualized in Fig.~\ref{fig:Model}. 

The calculated Dzyaloshinskii-Moriya interaction between nearest neighbors ($|{\bf D}_{01}|$ = 0.7 meV) is about two times larger than the resulting isotropic exchange interaction,  $J_{ij}=J^{kin}_{ij}-J^{F}_{ij}$ = -0.31 meV, which is ferromagnetic. This suggests Sn/SiC(0001) to be a physical realization of the system with dominant Dzyaloshinskii-Moriya interactions.
The model also contains symmetric anisotropic interactions expressed as a $\Gamma$ tensor, which defines the easy axis for each pair of magnetic moments. The principal axis of $\Gamma$ coincides with the direction of the corresponding Dzyaloshinskii-Moriya interaction, as follows from the single-band origin \cite{Aharony} of the magnetic interactions in Sn/SiC(0001). 

\begin{figure}[b]
\includegraphics[width=0.50\textwidth]{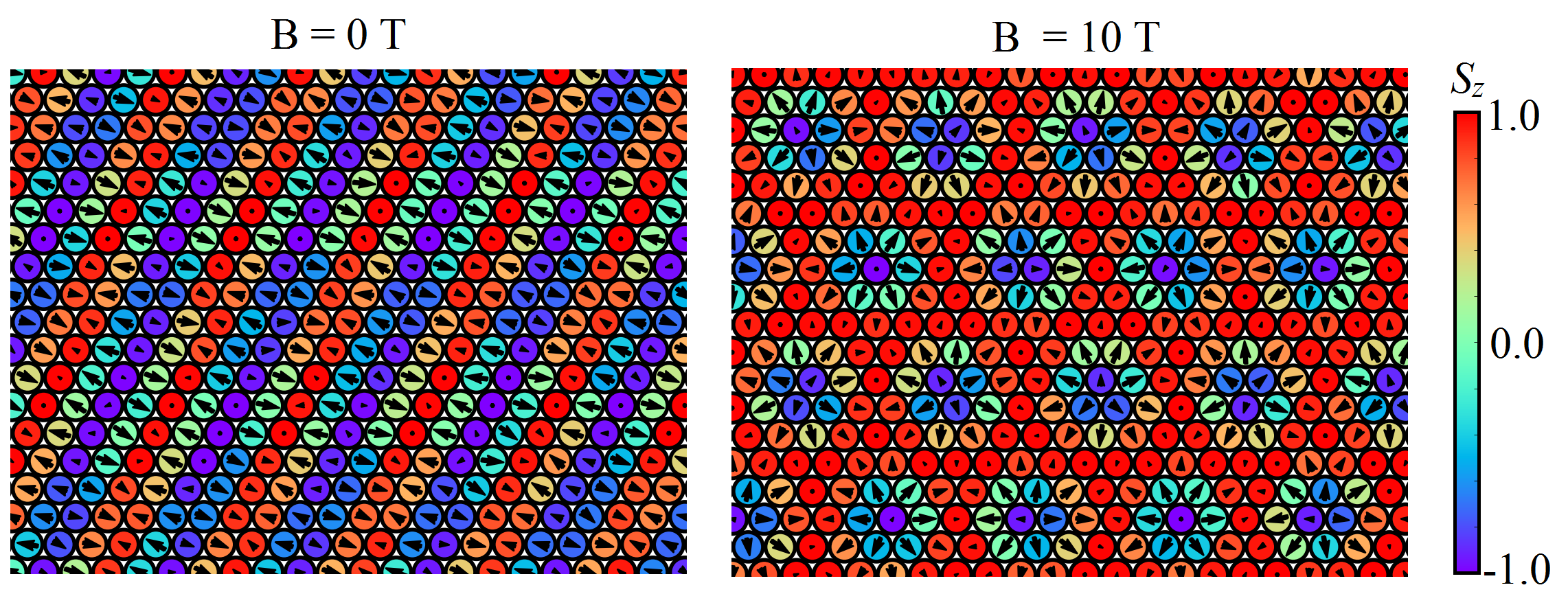}
\caption{ Snapshots of the Sn/SiC(0001) spin texture  obtained within Monte Carlo approach for different values of external magnetic field. The simulation temperature is 36 mK.}
\label{fig:MC}
\end{figure} 

\begin{figure*}[!t]
\includegraphics[width=0.95\textwidth]{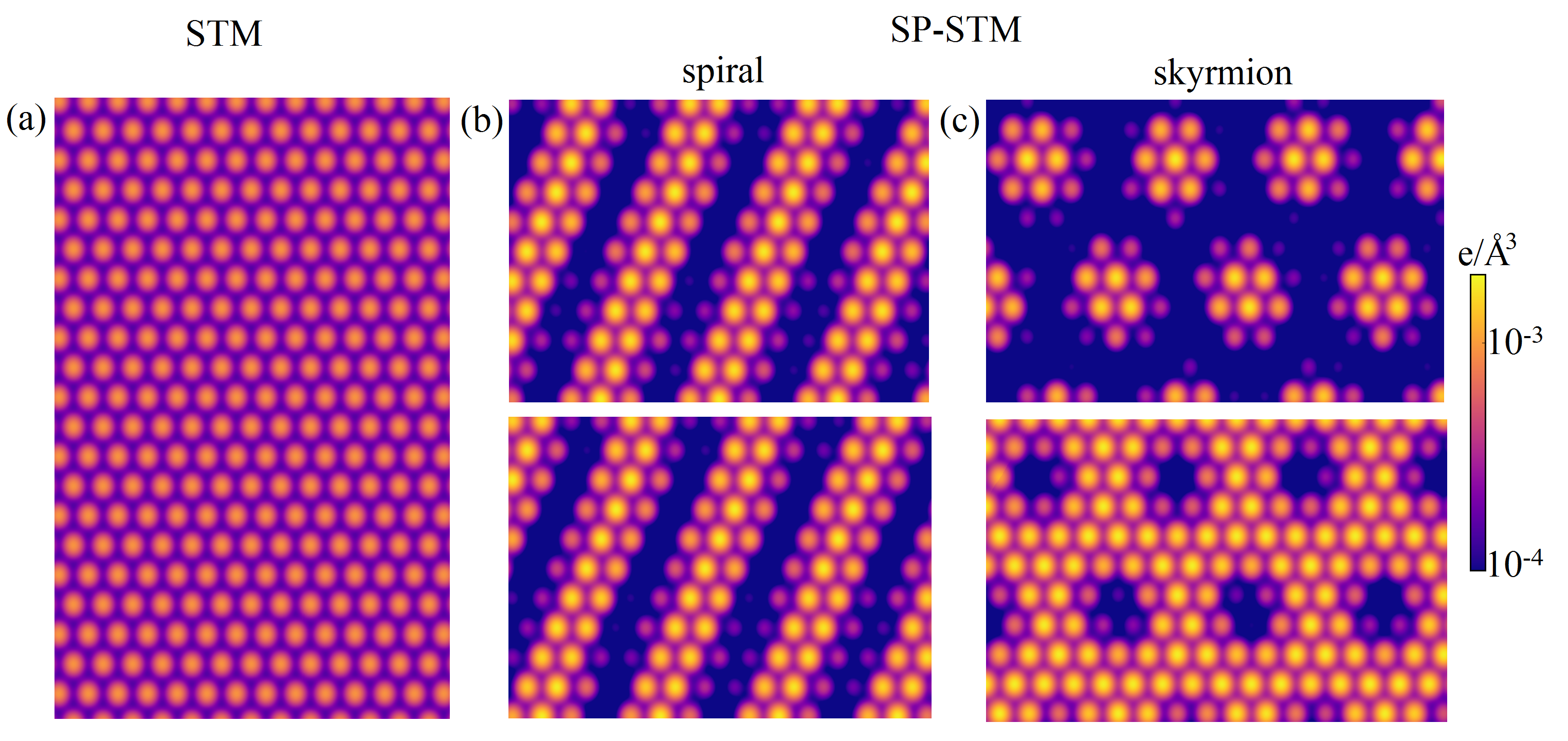}
\caption{STM images of Sn/SiC(0001) simulated by using the Tersoff-Hamann approach. (a) and (b,c) correspond to the simulations with the non-magnetic and magnetic tip, respectively. Spin-polarized images (b) and (c) were obtained by using the Monte Carlo  magnetic configurations presented in Fig.~\ref{fig:MC}. Top and bottom panels correspond to the tip's (0,0,$-$1) and (0,0,1) polarizations, respectively.  The distance between the tip and the surface is 4 \AA. Bias voltage is $-$2 V.  }
\label{fig:SP_STM}
\end{figure*}

{\it Nanoskyrmion state.} ---To explore the properties of the strongly anisotropic spin Hamiltonian constructed for Sn/SiC(0001), we perform classical Monte Carlo simulations at finite temperatures and magnetic fields. Such calculations demonstrate the existence of characteristic spin spiral and skyrmion states at realistic magnetic fields and temperatures (see Fig.~\ref{fig:MC}). The lowest critical magnetic field needed for the skyrmion lattice state formation is about 7 T. In turn, the temperature maximum for the skyrmion phase in Sn/SiC(0001) is about 1 K, which is also within the experimental range~\cite{Khajetoorians}. The dominant character of DMI makes the formation of compact skyrmions possible, with the radius of a few lattice constants.

To stimulate further experimental studies, we model the spin-polarized STM spectra within the Tersoff-Hamann theory~\cite{Tersoff,Blugel}. The results are shown in Fig.~\ref{fig:SP_STM}.
At realistic experimental parameters, we observe a striking contrast in the STM images between the patterns corresponding to nonmagnetic, spin spiral, and skyrmion states. Therefore, the distinct magnetic phases can be distinguished by means of available experimental instruments.

One important aspect for detecting the magnetic patterns in real SP-STM experiments is the existence of magnetic anisotropy, which couples spin to the lattice \cite{Loth}, and produces the spin-polarized contrast in STM images. Particularly, the anisotropy must be large enough to prevent temperature fluctuations of individual magnetic moments. While in our $S=1/2$ case there is no single-ion anisotropy, the intersite symmetric anisotropic exchange exists. To quantify the effect, we use the molecular field theory \cite{Mn12}, and summarize all the pair symmetric anisotropic exchange interactions for a given spin assuming that the spins are collinear, $E_{anis} = 2(\sum_{j} \Gamma_{0j}^{xx} - \sum_{j} \Gamma_{0j}^{zz}) |S|^2$. The obtained value of $E_{anis}$ is equal to 3.3 K. Our estimate for Sn/SiC(0001) is much smaller than that for surface nanosystems, where $S>\frac{1}{2}$ allowing for single-ion anisotropy~\cite{Loth}. 
We expect a strong enhancement of the intersite anisotropy by replacing Sn with heavier group IV elements. For instance, in the case of Pb/Si(111)~\cite{Danis} the anisotropy energy can be estimated to be around 32~K. 

{\it Conclusion.} ---We have explored magnetic properties of the Sn/SiC(0001) system on the basis of a generalized Hubbard model taking into account spin-orbit coupling, non-local Coulomb and exchange correlations. We found that Sn/SiC(0001) falls into an unusual regime of magnetic parameters, allowing for the existence of practically interesting phenomena, including spin spiral and skyrmion phases. Systematic theoretical characterization of Sn/SiC(0001) allowed us to suggest realistic experimental conditions, at which spin-polarized STM measurements can be performed to detect these phases.

{\it Acknowledgments.} --- This work was supported by the Russian Science Foundation, Grant No. 17-72-20041.

\pagebreak
\widetext
\begin{center}
\textbf{\large Supplemental Material: Nanoskyrmion engineering with $sp$-electron materials: \\ Sn monolayer on SiC(0001) surface}
\end{center}
\setcounter{equation}{0}
\setcounter{figure}{0}
\setcounter{table}{0}
\setcounter{page}{1}
\makeatletter
\renewcommand{\theequation}{S\arabic{equation}}
\renewcommand{\thefigure}{S\arabic{figure}}
\renewcommand{\bibnumfmt}[1]{[S#1]}
\renewcommand{\citenumfont}[1]{S#1}

\section{\label{sec:DFT} Details of DFT and DFT+SO calculations}
Electronic properties of the Sn/Si(0001) system were simulated within generalized gradient approximation (GGA) using the Perdew-Burke-Ernzerhof (PBE) exchange-correlation functional \cite{S_PBE} as implemented in the {\sc Quantum Espresso}~\cite{S_espresso} and Vienna ab-initio simulation package ({\sc VASP})~\cite{S_VASP,S_VASP_1}  with plane-wave basis set. The calculation parameters are as follows. The energy cutoff equals to 500 eV and the energy convergence criteria is 10$^{-8}$ eV. For the Brillouin zone integration a 20$\times$20$\times$1 Monkhorst-Pack mesh was used. A vacuum space of 16~\AA~between unit cell replicas in the vertical $z$ direction was introduced.  The optimized atomic positions and the band structure of the nonmagnetic ground state are in good agreement with previous works~\cite{S_Glass,S_Cho}. In particular, the corresponding Sn-Si and C-C distances are 2.05~\AA~and 0.18~\AA, respectively. The distance between nearest neighbor adatoms is 5.32~\AA, which is smaller than that for the Sn/Si(111) system, 6.61~\AA.

\begin{figure}[!h]
\includegraphics[width=0.7\textwidth]{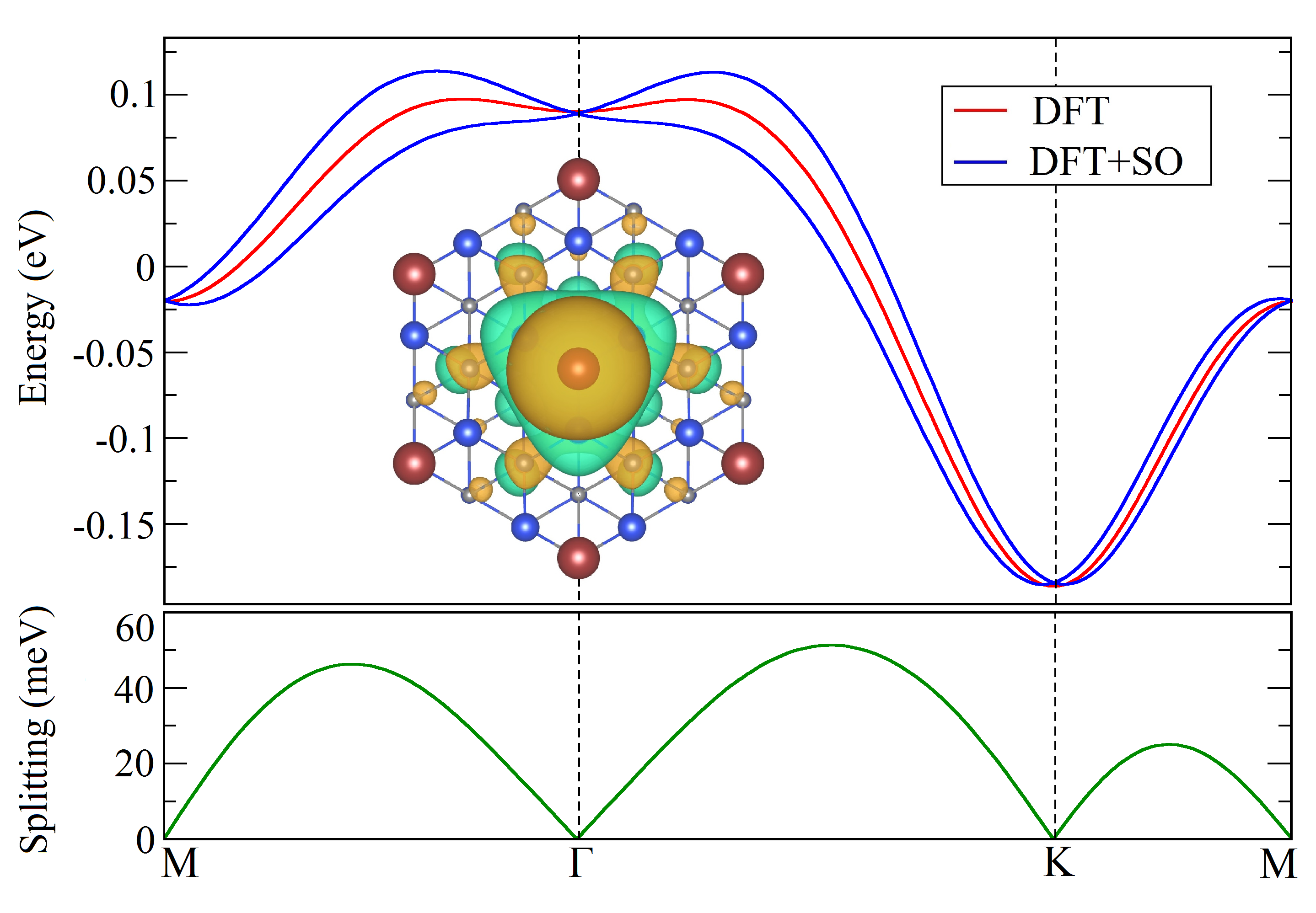}
\caption{ (Top) Band structures as obtained from DFT and DFT+SO calculations. The inset shows the Wannier function (top view) representing the band at the Fermi level. (Bottom) Energy splitting of the band in DFT+SO calculations due to spin-orbit coupling.}
\label{fig:BANDS_SO}
\end{figure}

Fig.~\ref{fig:BANDS_SO} shows the calculated electronic spectrum of the Sn/Si(0001) system. The width of the energy band at the Fermi level is 0.29 eV, which is smaller than the corresponding value of 0.45~eV in Sn/Si(111). Taking into account spin-orbit coupling leads to the energy splitting, which is in agreement with Ref.~\onlinecite{S_Glass}. 

To construct the low-energy tight-binding model in the Wannier functions basis, we use the procedure of maximal localization \cite{S_Marzari,S_Marzari2} implemented in {\sc wannier90} package~\cite{S_wannier90}. The resulting Wannier function is well localized with the corresponding spread of 7.9~\AA $^2$, which is about twice as small as in Sn/Si(111), 16.8~\AA $^2$~\cite{S_Danis}. This difference can be explained by the fact that the Wannier function in Sn/SiC(0001) is less hybridized with the slab compared to Sn/Si(111). A significant delocalization of the electronic density (Fig.~\ref{fig:BANDS_SO}) suggests that magnetic properties cannot be reliably described only by using the atomic wave functions.

\section{\label{sec:Low-energy models }Construction of the low-energy model}
{\it Electronic model.}
In order to study the electronic and magnetic properties of the Sn/SiC(0001) system, we use an effective electronic model constructed in the Wannier functions basis taking into account spin-orbit coupling and electronic correlations: 

\begin{eqnarray}
\hat{\cal H}=\sum_{i j,\sigma\sigma'}t_{ij}^{\sigma\sigma'}\hat{a}_{i \sigma}^{+}\hat{a}_{j \sigma'} 
+ \frac{1}{2}\sum_{i j k l,\sigma\sigma'} K_{ijkl}\,\hat{a}_{i \sigma}^{+}\hat{a}_{j \sigma'}^{+}\hat{a}^{}_{l \sigma'}\hat{a}^{}_{k \sigma},
\label{eq:electron_ham}
\end{eqnarray}
where $i(j)$ and $\sigma (\sigma')$ are site and spin indices; $t_{ij}^{\sigma \sigma'}$ is the element of the hopping matrices with spin--orbit coupling;  $K_{ijkl}$  represents the element of the Coulomb interaction matrix. To define the model parameters, we used the state-of-the-art approaches based on the Wannier functions formalism, which are described below.

{\it Coulomb interactions.}
In the static limit ($\omega$ = 0), the bare value of the four-element Coulomb integral in the basis of Wannier functions $W_i(\boldsymbol{r})$  can be defined as:
\begin{equation}
K_{ijkl} = \int  \frac{ W^{*}_i(\boldsymbol{r}) W_k(\boldsymbol{r})  W^{*}_j(\boldsymbol{r}') W_l(\boldsymbol{r}') }{|\boldsymbol{r}-\boldsymbol{r}'|}  d\boldsymbol{r} d\boldsymbol{r}'.
\label{Coulomb_bare}
\end{equation}
Here, we consider the on-site and inter-site Coulomb  interactions, $V_{ii}= K_{iiii}$ and $V_{ij} = K_{ijij}$, respectively, as well as the nonlocal direct exchange interaction, $J^{F}_{ij} = K_{ijji}$. To capture the effects of screening, we adopt a numerical procedure previously applied to $sp$ systems \cite{S_Mazurenko,S_Rudenko}. Particularly, we use the constrained random phase approximation (RPA)~\cite{S_RPA}:
\begin{equation}
\kappa (\boldsymbol{q}) = [1 - K(\boldsymbol{q})P(\boldsymbol{q})]^{-1}K(\boldsymbol{q}),
\label{Ploarization}
\end{equation}
where $P(\boldsymbol{q})$ is the static single-particle polarizability function: 

\begin{equation}
P(\boldsymbol{q}) = \frac{1}{\Omega} \sum_{m \boldsymbol{k}}^{occ} \sum_{n \boldsymbol{k}'}^{unocc} \frac{| \langle  \Phi_{m \boldsymbol{k}} | e^{-i \boldsymbol{qr} } | \Phi_{n \boldsymbol{k}'} \rangle |^2 }    {\epsilon_{m \boldsymbol{k}} - \epsilon_{n \boldsymbol{k}'} +i \eta}.
\label{Polarization_RPA}
\end{equation}
In Eq.~\ref{Polarization_RPA}, $\boldsymbol{k}' = \boldsymbol{k} + \boldsymbol{q}$, summation runs over the Brillouin
zone including transitions from occupied $m$ to unoccupied $n$ states. Account of all possible transitions from occupied to empty states gives the fully screened value of the corresponding Coulomb integrals. At the same time, excluding ``self-screening'' transitions for the band at the Fermi level gives a partially screened value of $\kappa (\boldsymbol{q})$~\cite{S_silke2}. Here, $\Omega$ is the volume of the unit cell; $\epsilon_{m \boldsymbol{k}}$ and $\Phi_{m \boldsymbol{k}}$ stand for the eigenvalues and eigenvectors of the full DFT Hamiltonian, respectively; $\eta$ is a numerical smearing parameter chosen to be 10 meV. The bare $V$, partially screened $U$ and fully screened $W$ values of the on-site and inter-site Coulomb interactions were evaluated for Sn/SiC(0001), and are presented in Table~\ref{tab:Coulomb}. 

\begin{table}[!h]
\centering
\caption [Bset]{On-site (00) and inter-site (01) values (in eV) of the bare $V$, partially screened $U$ and fully screened $W$ Coulomb interactions, as obtained by using cRPA. See Fig.~\ref{fig:Model} for further details on the interaction paths. }
\begin{ruledtabular}
\begin {tabular}{c|cccc}
  
 path & $V$ & $U$ & $W$  \\
\hline
00 &  5.89 & 1.98 & 0.42 \\
01 &  2.45 & 0.80 & 0.34 \\
\end {tabular}
\end{ruledtabular}
\label{tab:Coulomb}
\end {table}

\begin{figure}[!h]
\includegraphics[width=0.7\textwidth]{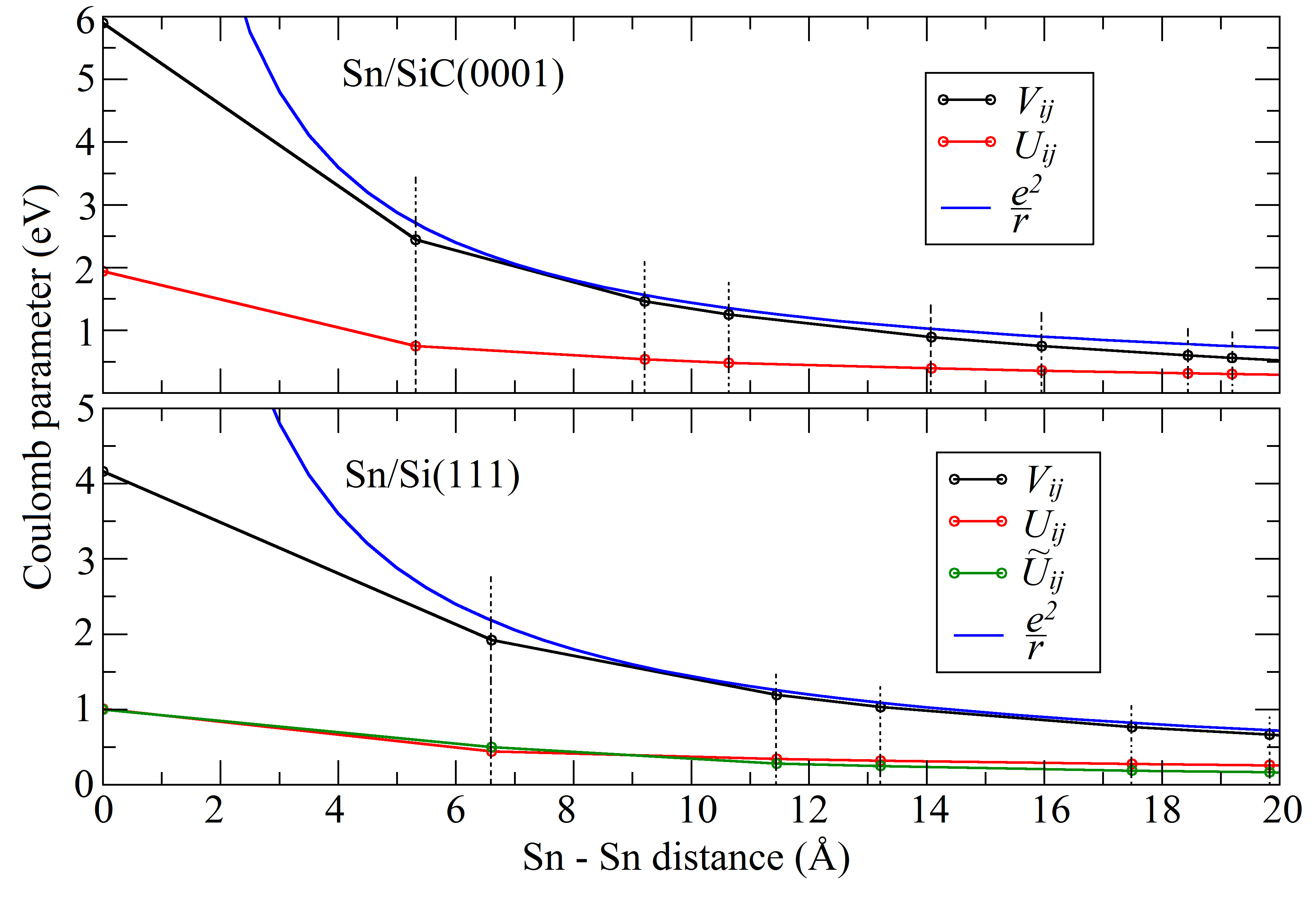}
\caption{Calculated Coulomb interactions for Sn/SiC(0001) (top panel) and Sn/Si(111) (bottom panel) systems. The values denoted with tilde (green line) are taken from work \cite{S_silke1}.}
\label{fig:Coulomb}
\end{figure}

For comparison, we calculated the bare and screened values of the Coulomb interactions in Sn/Si(111). The obtained results are in good agreement with reported previously values~\cite{S_silke1}. For both systems, the bare parameters at long distances follow the classical Coulomb law $e^2/r$. For nearest and next-nearest neighbors, the deviation from the $e^2/r$ law can be explained by the fact that a particular shape of the constructed Wannier functions plays a dominant role when calculating their couplings. At the same time, at long distances the corresponding Wannier functions can be considered as point-like objects and their particular form is not important. Dependence of the Coulomb interactions on screening is almost the same for both the on-site and inter-site interactions (Fig.~\ref{fig:Coulomb}).  

Generally, the direct exchange interaction parameter $J^{F}_{ij}$ is relatively small compared to the on-site and inter-site Coulomb integrals. As a consequence, the application of the cRPA scheme requires an extremely accurate integration over the Brillouin zone in order to calculate the polarizability function (Eq.~\ref{Polarization_RPA}), which is computationally challenging. We estimated the bare value of the direct exchange interaction,  $J^{F \, (bare)}_{01}$ = 5.20 meV, that defines the upper boundary for this coupling. To take into account the effects of screening, we used the renormalization factor of 3 obtained for the on-site and inter-site Coulomb interactions. Thus, one obtains $J^{F \, (screened)}_{01} = J^{F \, (bare)}_{01}/3$ = 1.73 meV. This value was used to define the total exchange interaction.   

{\it Hopping integrals.}
The calculated hopping parameters $t_{ij}^{\sigma \sigma'}$ in the Wannier functions basis are shown in Table~\ref{tab:hoppings}. Their diagonal parts are in excellent agreement with previously reported values obtained without spin-orbit coupling~\cite{S_Glass}.  As will be shown below, significant non-diagonal elements of the hopping matrix in spin space indicate that there are strong anisotropic magnetic excitations. 

\begin{table}[!h]
\centering
\caption [Bset]{Hopping integrals (in meV) between nearest and next-nearest neighbors as obtained from DFT+SO calculations. See Fig.~\ref{fig:Model} for further details on the interaction paths.}
\begin{ruledtabular}
\begin {tabular}{cccc}
   & $t_{01}$ & $t_{02}$ & \\
\hline
  & $\left(\begin{array}{cc} 27.94  &  7.39   \\ -7.39 & 27.94  \end{array} \right)$ &  $\left( \begin{array}{cc} -14.36 + 0.12 i &  -1.50 i \\  -1.50 i & -14.36 - 0.12 i \end{array}\right)$\\
\end {tabular}
\end{ruledtabular}
\label{tab:hoppings}
\end {table}

\section{Hartree-Fock solution}

The Hartree-Fock method provides the simplest approximation to the many-body problem \citep{S_Solovyev_HF}. Within this approach, the electronic model given by Eq.~\ref{eq:electron_ham} is solved as:

\begin{equation}
\left(\hat{t}_{\boldsymbol{k}}+\hat{\mathcal{V}}^{H}_{\boldsymbol{k}} + \hat{\mathcal{J}}^{H}_{\boldsymbol{k}} \right)|\varphi_{\boldsymbol{k}}\rangle=\varepsilon_{\boldsymbol{k}}|\varphi_{\boldsymbol{k}}\rangle,
\label{hf}
\end{equation}
\noindent where $\hat{t}_{\boldsymbol{k}}$ is the Fourier transform of the hopping parameters $\hat{t}_{ij}$, $\hat{\mathcal{V}}^{H}_{\boldsymbol{k}}$ and $\hat{\mathcal{J}}^{H}_{\boldsymbol{k}}$ are the Hartree-Fock potentials describing the Coulomb and direct exchange interactions, respectively, $\varepsilon_{\boldsymbol{k}}$ and $|\varphi_{\boldsymbol{k}}\rangle$ are the corresponding eigenvalues and eigenvectors in a given basis; a self-consistent solution of Eq.~(\ref{hf}) is achieved with respect to the density matrix:
\begin{equation}
\hat{n}=\sum\limits_{\boldsymbol{k}}|\varphi_{\boldsymbol{k}}\rangle \langle\varphi_{\boldsymbol{k}}|.
\end{equation}

\begin{figure}[!h]
\includegraphics[width=0.7\textwidth]{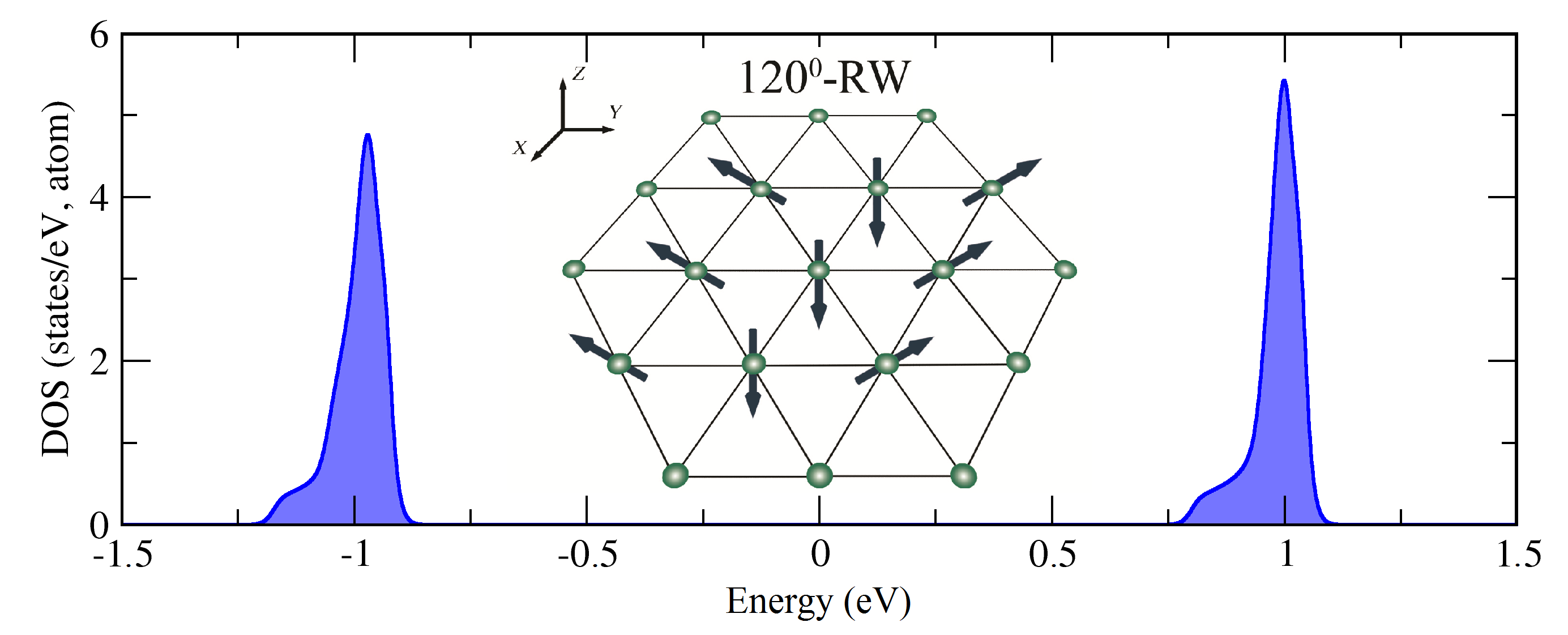}
\caption{  Density of states (DOS) of Sn/SiC(0001) as obtained by using the Hartree-Fock approximation for the full model (Eq.~\ref{eq:electron_ham}). The ground-state magnetic order is shown in the inset.}
\label{fig:HF_DOS}
\end{figure} 

In our study, we simulated four magnetic structures: ferromagnetic (FM), 120$^\circ$-N\'eel (or 120$^\circ$-AFM), collinear row-wise (cRW or cAFM) and  120$^\circ$ row-wise (120$^\circ$-RW) structures for 3 $\times$ 3 unit cell. Due to strong spin-orbit couping, the FM and cRW configurations are found to be unstable. Having compared the energies of the noncollinear solutions, we conclude that the 120$^\circ$-RW state corresponds to the ground state of the Sn/SiC(0001) system (Fig.~\ref{fig:HF_DOS}). The calculated density of states (DOS) has the insulating gap of 2.0 eV, which is in excellent agreement with the results of photoemission experiment~\citep{S_Glass}. 

\section{Construction of the spin model}
To explore the excited magnetic states of the Sn/SiC(0001) system, we construct the spin model, given by Eq.~\ref{spinham} in the main text. The parameters of this model are estimated at the level of the superexchange theory~\cite{S_Anderson}. Particularly, the isotropic term consists of the antiferromagnetic kinetic part $J_{ij}^{kin}$ and the screened direct exchange interaction $J_{ij}^{F}$:
\begin{eqnarray}
J_{ij} =   \frac{1}{\widetilde U} {\rm Tr_{\sigma}} \{ \hat t_{ji} \hat t_{ij} \} -  J^{F}_{ij},
\label{exch}
\end{eqnarray}
where $\hat t_{ij}$ is the hopping integral with spin-orbit coupling, the effective local Coulomb interaction, $\widetilde U$ is estimated as $\widetilde U = U - V_{ij}$. The non-local Coulomb interaction plays a significant role for nearest neighbor interactions only. For the long-range interactions, its value is further reduced due to the exponential decay of the Wannier functions overlap and screening effects. The calculation gives $J_{01} = 1.42 - 1.73 = -0.31$ meV, whereas for the next-nearest interaction $J_{02}$ = 0.21 meV.

As for the anisotropic part of the magnetic model, the antisymmetric Dzyaloshinskii-Moriya (DM) and symmetric anisotropic exchange interactions are given by~\cite{S_Aharony1}:
\begin{equation}
\boldsymbol{D}_{ij} = \frac{i}{2\widetilde{U}}[{\rm Tr}(\hat {t}_{ij}){\rm Tr}(\hat{t}_{ji}\boldsymbol{\sigma})-{\rm Tr}(\hat{t}_{ji}){\rm Tr}(\hat{t}_{ij}\boldsymbol{\sigma})],
\label{eq:DM-vector}
\end{equation}
\begin{equation}
\overset{\leftrightarrow}{\Gamma}_{ij} = \frac{1}{2\widetilde{U}}[{\rm Tr}(\hat{t}_{ji}\boldsymbol{\sigma})\otimes {\rm Tr}(\hat{t}_{ij}\boldsymbol{\sigma})+{\rm Tr}(\hat{t}_{ij}\boldsymbol{\sigma})\otimes {\rm Tr}(\hat{t}_{ji}\boldsymbol{\sigma})],
\label{eq:Gamma-matrix}
\end{equation}
where $\boldsymbol{\sigma}$ are the Pauli matrices. The DMI vectors depicted in Fig.~2 paths of interactions are the following:  ${\bf D}_{01} =  (0, 0.70, 0)$, ${\bf D}_{02} = (0.043,  0, -0.003)$ meV. In accordance with the $C_{3v}$ symmetry, the nearest neighbor DMI vectors lie in $xy$-plane and are perpendicular to the corresponding bond, whereas the next-nearest neighbor DMI vectors alternate their $z$ component. 

The calculated symmetric anisotropic exchange interaction tensor has the following form (in meV):
\begin{eqnarray}
\Gamma_{01} = \left( \begin{array}{ccc} 0.00  &  0.00 & 0.00  \\ 0.00  & 0.19 & 0.00 \\ 0.00 & 0.00 & 0.00  \end{array}\right).
\end{eqnarray}
The tensors for the rest of the bonds can be obtained by 60$^{\circ}-$rotations around the z axis. 

\section{\label{sec:Magnetic characteristic}Analysis of the DFT magnetic moments and total energies }
In order to elucidate the magnetic structure and capture the strong delocalization effects, we performed the calculations by using an all-electron full-potential linearized augmented-plane wave method (FP-LAPW) implemented in the ELK code \cite{S_ELK}. Without spin-orbit coupling a significant interstitial contribution to the band at the Fermi energy (Fig.~\ref{fig:BANDS_SO}) was revealed. It is interesting to note that the magnetic moment has a similar composition. Spin-polarized calculations for the ferromagnetic state give 1 $\mu_B$, which contains only $\sim$50\% of the total moment at muffin-tin spheres, while the other half belongs to the interstitial area (Table.~\ref{tab:Magnetic_moments}). 

\begin{table}[!h]
\centering
\caption [Bset]{The total value and partial contributions to the magnetic moment obtained for the ferromagnetic ordering by using different first-principles methods (in $\mu_B$). Notation $MT$ and $I$ stands for sum of the moments in muffin-tin spheres and interstitial contribution, respectively.}
\label {basisset}
\begin{ruledtabular}
\begin{tabular}{lccc}
   & Quantum Espresso & VASP & ELK  \\
  \hline
 M$_{Sn}$  & 0.15 & 0.23 & 0.21   \\
 M$_{MT}$  & 0.35 & 0.52 & 0.51   \\
 M$_{Tot}$ & 1.00 & 1.00 & 0.96   \\
 M$_{I}$   &  -   &   -  & 0.45   \\

\end{tabular}
\end{ruledtabular}
\label{tab:Magnetic_moments}
\end {table}

\begin{figure}[!h]
\includegraphics[width=0.7\textwidth]{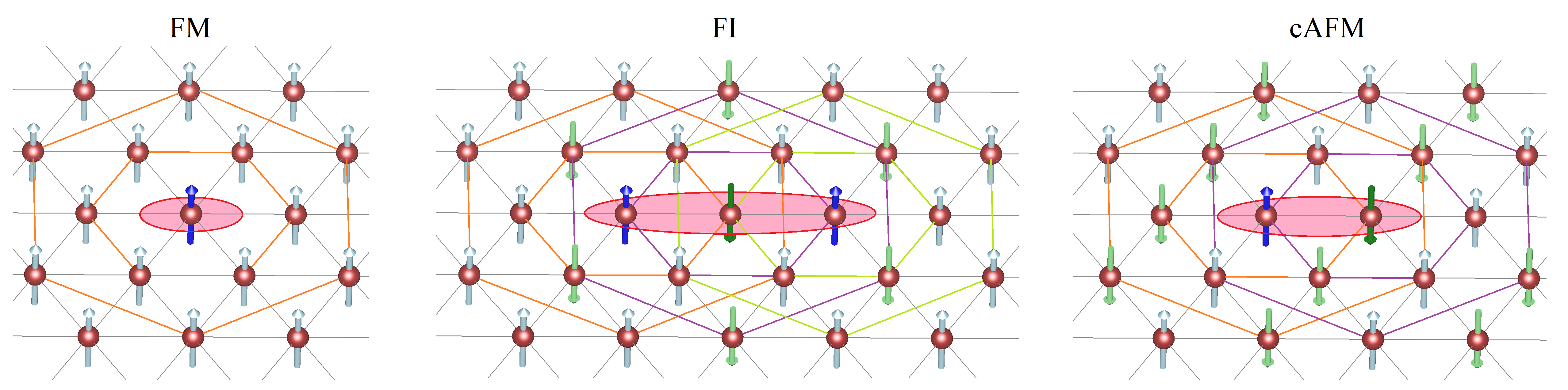}
\caption{ Ferromagnetic (FM), collinear antiferromagnetic (cAFM) and ferrimagnetic (FI) configurations simulated in previous works \cite{S_Glass, S_Cho_Si_111}. Arrows represent the directions of local magnetic moments. Red circles denote the magnetic unit cell.}
\label{fig:cAFM}
\end{figure} 

In the previous calculations~\cite{S_Glass, S_Cho}, the authors considered ferromagnetic (FM), ferrimagnetic (FI) and two antiferromagnetic magnetic configurations: collinear (cAFM) and non-collinear 120$^\circ$ (120$^\circ$-AFM) ordering (Fig.~\ref{fig:cAFM}). It was reported that cAFM corresponds to the minimum of the total energy in Sn/SiC(0001). By using the calculated total energies of different magnetic configurations it is possible to estimate the individual magnetic interactions, between nearest and neighbors, $J_{01}$ and $J_{02}$, respectively. For that purpose, we consider the Heisenberg spin Hamiltonian $ \mathcal {H} = \sum_{ij} J_{ij} {\bf S}_{i} {\bf S}_{j}$. For convenience, a $3 \times 3$ unit cell is used. In this case, the energies for these magnetic configurations can be written in the following form:
\begin{equation}
 \begin{aligned}
    3E_{FM} = (18J_{01} + 18J_{02}) \times \frac{1}{4} \\        
    3E_{cAFM} = -(6J_{01} + 6J_{02}) \times \frac{1}{4} \\
    3E_{FI} = - (6J_{01} - 18J_{02}) \times \frac{1}{4} 
  \end{aligned}
\label{eq:Energies}
\end{equation}
Here, the energies $E_{FM}$, $E_{cAFM}$ and $E_{FI}$ are given per $\sqrt{3} \times \sqrt{3}$ unit cell. From these equations one can derive the expressions for nearest and next-nearest neighbors isotropic exchange interactions:

\begin{equation}
 \begin{aligned}
     J_{01} = \frac{1}{2} (E_{FM} - E_{FI}) \\
     J_{02} = \frac{1}{2} (E_{FI} - E_{cAFM}) \\
  \end{aligned}
\label{eq:J_calculation}
\end{equation}
The resulting interactions are given in Table~\ref{tab:Energies} in the main text.

\begin{table}[!h]
\centering
\caption [Bset]{Energies and corresponding values of isotropic exchange interactions (in meV) for Sn/SiC(0001) and Sn/Si(111) estimated by using the results of previous DFT calculations with different exchange-correlation (xc) functionals. Energies are given in meV per $\sqrt{3} \times \sqrt{3}$ unit cell.  Values in brackets are those evaluated with superexchange theory in this work and in Ref.~\onlinecite{S_Danis}. }
\label {basisset}
\begin{ruledtabular}
\begin{tabular}{ccccccc}
 &  xc-type &  $E_{FM}$ & $E_{cAFM}$ & $E_{FI}$ & $J_{01}$ & $J_{02}$ \\
  \hline
  {Sn/SiC(0001)} &  & & & & & \\
  \\
Glass {\it et al.}, Ref.~\onlinecite{S_Glass} & LDA & 3.03 & $-$2.93 & $-$2.03 &  2.53\,(1.42)  & 0.45\,(0.21) \\
   \hline
   \\
  {Sn/Si(111)} &  & & & & & \\
  \\
Cho {\it et al.}, Ref.~\onlinecite{S_Cho_Si_111} & hybrid & $-$47.9 & $-$65.1 & $-$64.2 &  8.15\,(7.69) & 0.45 \,(0.73) \\
 
\end{tabular}
\end{ruledtabular}
\label{tab:Energies}
\end {table}

\section{\label{sec:Monte Carlo simulation}Monte Carlo simulation}
Monte Carlo simulations were performed by using the heat-bath method combined with overrelaxation~\cite{S_Binder}. The corresponding model parameters are given up to next-nearest neighbors. In these calculations, supercells of various sizes from $N=96\times96$ to $150\times150$ spins with periodic boundary conditions were used, and a single run contained $(0.5-2.0)\cdot 10^{6}$ Monte Carlo steps. For initial relaxation the system is gradually cooled down from higher temperatures.

\section{Scanning tunneling microscopy (STM) simulations }
Spin-polarized scanning tunneling microscopy technique \cite{S_Wiesendanger} is a powerful tool for exploring distinct spin textures on a surface, such as spin spirals or skyrmions. 
Theoretical background for such experiments was developed in Refs.~\cite{S_Fransson,S_Blugel}. According to these studies, the tunneling current can be written in the following form: 
\begin{eqnarray}
I(\boldsymbol{r},eV) \sim \int [f(\epsilon)-f(\epsilon-eV)]\times[n(\epsilon-eV)N(\boldsymbol{r},\epsilon)+  \boldsymbol{m}(\epsilon-eV)\boldsymbol{M}(\boldsymbol{r},\epsilon)] d \epsilon,
\label{eq:STM}
\end{eqnarray}
where $n(\epsilon)$ stands for the tip's DOS, $N(\boldsymbol{r},\epsilon)$ is the DOS of the surface. The magnetic part of the tunneling current contains the tip's and surface magnetization densities, $\boldsymbol{m}(\epsilon)$ and $\boldsymbol{M}(\boldsymbol{r},\epsilon)$, respectively. $V$ denotes the bias applied to the sample. To define the electronic and magnetization densities, we employed the calculated Wannier functions, $W(\boldsymbol{r})$. It allows us to take into account the delocalization of the electronic state and magnetic moment. Thus, one has $N(\boldsymbol{r},\epsilon) = |W(\boldsymbol{r})|^2 N_{loc}(\epsilon)$ and $\boldsymbol{M}(\boldsymbol{r},\epsilon) = |W(\boldsymbol{r})|^2 \boldsymbol{M}_{loc} N_{loc}(\epsilon)$, where  $N_{loc}(\epsilon)$ and $\boldsymbol{M}_{loc}$ is the local DOS and magnetic moment of adatom obtained from DFT calculations and Monte Carlo simulations, respectively.  Practically, we assumed that the tip's DOS $n(\epsilon)$  and magnetization $\boldsymbol{m}(\epsilon)$ do not depend on energy.

\end{document}